%Paper: hep-th/9307073
%From: Cobi Sonnenschein <COBI@taunivm.tau.ac.il>
%Date: Fri, 09 Jul 93 23:21:04 IST

%macropackage=phyzzx
\overfullrule=0pt
\twelvepoint

\def\harveqn#1#2{$$#2\eqn#1$$}
\def\ack{\centerline{{\bf Acknowledgement}}\medskip\parindent=40pt}
\def\VW{$V$ and $W^3$}
\def\pa {\partial}
\def\nq{ n+{\alpha\over q}}
\def\oq{ $(1,q)$\ }
\def\pka{P_{k,\alpha}}
% journal definitions
%
\def\cmp#1{{\it Comm. Math. Phys.} {\bf #1}}

\def\pl#1{{\it Phys. Lett.} {\bf #1B}}
\def\prl#1{{\it Phys. Rev. Lett.} {\bf #1}}

\def\np#1{{\it Nucl. Phys.} {\bf B#1}}

\def\jmath#1{{\it J. Math. Phys.} {\bf #1}}
\def\mpl#1{{\it Mod. Phys. Lett.}{\bf A#1}}
\REF\DPGZ{ For a review of the subject see for instance
 P. Di Francesco, P. Ginsparg, and
 J. Zinn-Justin,
  LA-UR-93-1722, SPhT/93-061, hep-th/9306153.}
\REF\DIJ{ R. Dijkgraff ``Intersection theory, integrable hierarchies and
topological field theory" IASSNS-HEP-91/91,\hfill\break
 and references therein.}
\REF\DDK{F. David \mpl {3}  (1988) 1651;\hfill\break
J. Distler and H. Kawai \np
{321} (1989) 509.}
 \REF\Pol{A. M. Polyakov, \mpl  {2}  (1987) 893.}
\REF\BKDSGM{D. Gross and A. A. Migdal \prl {64} (1990) 127;\break
M. Douglas and S. Shenker \np {335} (1990) 635;\break E. Brezin and V. A.
Kazakov, \pl {236} (1990) 144.}
\REF\Du{  M. Duglas, \pl {238} (1990) 176.}
\REF\DVV{R. Dijkgraaf, E. Verlinde, and H. Verlinde,
\np {348} (1991) 435;\np {352} (1991) 59.}
\REF\FKN{M. Fukuma, H. Kawai and R. Nakayama,
{\it Int. Jou. of Mod. Phys.} {\bf A6} (1991) 1385.}
\REF\SY{M. Spigelglas and S. Yankielowicz, \np{393} (1993) 301.
\hfill\break E. Witten,
\cmp 144 (1992) 189.}
\REF\wtrr{E. Witten, \np {340} (1990) 281.}
\REF\VV{ E. Verlinde, and H. Verlinde,
\np {348} (1991) 457.}
\REF\Kon{M. Kontsevich \cmp {147} (1992) 1.}
\REF\Morozov{ For a review  see for instance A. Morozov
``Integrability and Matrix Models" ITEP-M2/93, hep-th9303139. }
\REF\Go{J. Goeree, \np{358}(1991)737.}
\REF\Li{K.Li, \np{ 354} (1991) 725-739.}
\REF\AMS{K. Aoki, D. Montano, J. Sonnenschein
{\it Int. Jou. of Mod. Phys.} {\bf A7} (1992)1755 .}
\REF\MR{  D. Montano and G. Rivlis,
``Solving  topological  2-D  quantum gravity using Ward identities"
 UCB-PTH-92-30 hep-th 9210106.}
\REF\MRR{  D. Montano and G. Rivlis, \np{360} (1991) 524.}
\REF\ABEL{Abel ,`` Oeuvers Comtletes", Cristiana, C. Grondhal 1893.}

%%%%%%%%%%%%%%%%%%%%%%%%%%%%%%%%%
\rightline{ TAUP 2073-93}
\titlepage
\title{Do we need the $W^{(n>3)}$ constraints to solve the $(1,q)$ models
coupled to 2D gravity?}
\author{Y.Lavi , J.Sonnenschein  }
\bigskip\centerline{School of Physics and Astronomy}
\centerline{Raymond and Beverly Sackler Faculty of Exact Sciences}
\centerline{Tel Aviv University}\centerline{Tel Aviv 69978, Israel}
\vskip .5in
\abstract
{We prove that all the  correlation functions in the $(1,q)$ models are
calculable  using only the Virasoro and the $W^{(3)}$ constraints.
This result is based on the invariance of correlators with respect
to  an interchange of
the order of the operators they contain.
 In terms of the  topological recursion relations,
 it means that only
two and three contacts and  the corresponding  degenerations
of the underlying surfaces  are
relevant. An algorithm to compute  correlators for
any $q$ and at any genus
is presented and  demonstrated through  some examples. On route to these
results, some interesting polynomial identities,
which are generalizations of Abel's identity,
were discovered. }
\eject

\section{Introduction}
Two dimensional (2D) gravitational models in general and topological ones in
particular attracted much attention in recent  years.\refmark\DPGZ\
This trend  was motivated mainly by the search for a
framework to investigate the non-perturbative behavior  of
string theory  and  to address fundamental
 questions of quantum gravity.    Mathematical questions  like the
topological properties of  certain moduli spaces  were additional motivations
for
 this research effort.\refmark\DIJ

The 2D gravitational models, which are often referred to as  non-critical
string models,  have been investigated in a variety of methods.  Among
them one  finds: (i)  The continuum formulations of  Liouville
theory,\refmark\DDK\
 and the world-sheet  light-cone  gauge.\refmark\Pol\  (ii)  The
discrete approach which started with the
 matrix models\refmark\BKDSGM\ and evolved into the  application of  KdV
flows.\refmark\Du\  (iii) The $W^{(n)}$ constraints on the partition
function.\refmark{\DVV,\FKN}
 (iv) The ${G\over G}$ topological coset approach.\refmark {\SY}\ (v) The
topological recursion relations approach.\refmark{\wtrr,\VV}
 (vi)  The Kontsevitch
integral formulation\refmark\Kon\ for the $(1,2)$ model and its
generalizations.    Since only $W^{(n)}$ constraints and the topological
recursion relations approach are directly related to our approach,  we
 summarize certain developments  only in these directions.

The KdV hierarchy\refmark\Morozov\ provides in principle a tool to
completely solve  the non-critical string models. In practice, however,  it is
not an easy task to solve the corresponding non-linear differential equations.
It was found that the KdV $\tau$ function, which determines the partition
function of  topological gravity , the topological  $(1,q=2)$ model,  had to
satisfy a set of constraints that obey half of a Virasoro
algebra.\refmark\DVV\
 Whereas  this set of constraints  is enough to
completely solve  the $q=2$ case, it is insufficient for  $q>2$.
A generalization  of the Virasoro constraints
to  a set of constraints associated
with the $W^{(3)}$ algebra produced a recursion relations
for the $q=3$ case\refmark{\Go,\Li}
which were  shown to be in accordance with the KdV solution.
Recursion relations based on  higher $W^{(q)}$ constraints were not written
down. Presumably  the cumbersome expressions found for the case of  two
primaries
 discouraged people to proceed in this direction

  The topological  recursion relation  approach was initiated by
Witten\refmark\wtrr\  for the   case of the topological gravity model which was
conjectured to be related to the one matrix model.
The main idea in this approach is that correlators  of topological models
should be determined  from contacts between the operators  and between
them  and possible   degenerations of
the underlying manifolds. By its nature it relies very little on the explicit
details of the field theory description.
  The introduction of the contact
algebra concept  combined with the requirement for invariance of
correlators
 under the
interchange of the order of the operators,
  led E. and H. Verlinde\refmark\VV\ to a full solution of the topological
gravity theory.
 This was  further developed  in ref. [ \AMS]  where
the contact algebra for the $(1,2)$ model was proven to be unique and the
role of multicontacts for higher $q$ was emphasized.
Further progress in this  program of analyzing  the general $(1,q)$    models
was made by
 D. Montano and G. Rivlis\refmark\MR (MR). They   provided
recursion relations for any correlator  of the $(1,q)$ model on the sphere
and for the case of  $q=3$ also for all higher genus.
Their approach included the  insertion of  a ``complete" set of states at each
degeneration of the surface and summing over all  the possible
degenerations.
 However, this prescription  does not
enable  one to deduce  a full solution  for $q>3$
 mainly due to the lack of a consistent
normal ordering technique to handle divergences that occur frequently at
higher genus.

In an attempt to generalize the  topological recursion relations of ref. [\MR]
and to  explicitly construct  $W^{(n>3)}$ constraints we discovered that in
fact   one can  compute correlators of $q>3$  and on higher genus
Riemann surfaces  only  with the aid of the Virasoro and $W^3$ constraints
(\VW).  Equivalently one can use the MR method with
 $q=3$,  namely,   only the two and three contacts.
Note that  for this particular value of $q$, it is  free
from any normal ordering problems.  These observations were later  translated
into   a proof based on induction  for every correlator  of the $q>3$ models.
The proof includes an algorithm for  the explicit computation of
any correlator at any genus.
In the usual  topological recursion relations  a given correlator  is
expressed in terms of a sum of correlators of less operators or correlators
at lower genus
or products of the latter.  The idea  behind our result is to go backward and
express a correlator of operators which cannot be manipulated  only by the
\VW constraints as a part of the recursion sum  of a correlator which include
in addition to the original
 operators  certain operators which are
descendants of the first two primaries.  For the latter correlator two
different
$W^3$ recursions are applied in such a way that   the coefficients in front
of the original correlator  are different.  Thus  by taking the difference of
the two sums one finds a recursion expression for the original correlator.
Obviously we use here the requirement that a correlator is independent of
the order of the operators it contains.  It is interesting to note that this
``physical" assumption translates into  the commutativity of derivatives with
respect to the coupling  in the KdV language.  This turns into a non-trivial
condition when one construct the  corresponding   $W^{(n)}$constraints.

   The paper is organized as follows.
In section 2   we  review the application of  $W_q$ constraints and
the topological procedure of ref. [\MR] to the computation of correlators in
the  $(1,q)$ models.  The relation to the KdV hierarchies is also briefly
discussed.  Section 3 is devoted to a  proof   that every
correlation function for the $q=4$ case  can be computed using  only the \VW
constraints.  The proof is based on  an induction algorithm both in the
number of operators in the  correlator and the genus of the underlying
Riemann surface.  In section 4 the proof is generalized to the general $q>
4$ models by invoking a further induction in the minimal primary degree
of the operators.
A summary  and  several open questions are presented in section 5.
 Appendix A  presents  certain explicit calculations
of correlators for $q=4\ $. These computations
demonstrate the usefullness of the
methods introduced in the proofs presented in sections  3.
In Appendix B  we write down a proof of the polynomial identities
which are used to derive the result for the general $(1,q)$ model.

\section{  Recursion relations from $W_q$ constraints and from the MR
topological procedure  }
The \oq  models are a class of topological models with $q-1$ primaries.
They are believed to be a description of the \oq minimal models coupled to
2D gravity.
The  correlation functions of these models can be determined by the \oq
KdV hierarchy. The basic differential operators  of  the KdV flows take the
form $P =D={\pa\over \pa x}$ and $Q =D^q+x$  where $x$ is the
cosmological  constant. The correlators of the theory are expectation values
of products of  primary operators and their descendants. These
operators are denoted by $P_{k,\alpha}$ where $\alpha=1,...,q-1$.  The
$k=0$ operators are the primaries and their descendants carry a positive
integer $k$.  There is a $U(1)$  ``ghost number"  charge associated with each
operator
$$gh(\pka) = [(k-1)q + (\alpha-1)]  \eqn\mishgh$$
such that non-trivial correlators $<\prod_i^n P_{k_i\alpha_i}>_g$
 at $x=0$  have to obey the  following  ``ghost number"
conservation  law
$$\sum_i^n gh(P_{k_i\alpha_i}) = 2(g-1)(1+q)\eqn\mishcgh$$

It was shown that  the $\tau$ function   of the KdV models is constraint by
a set of operators which furnish an anomaly free  half Virasoro
algebra.\refmark{\DVV,\FKN}.  These constraints completely determine all
the correlators of the $(1,2)$ model,  but are not enough to do a similar job
for models with $q>2$.  Since the $W_q$ algebra
 forms a  closed algebra which includes
 the Virasoro generators    it was proposed
that
    invoking $W_q$
constraints on the $\tau$ function would lead to a complete
derivation of  recursion relations for the  $(1,q)$ models.
Indeed, for the  case of two primaries, $q=3$, recursion
relations were  derived using this approach.\refmark{\Go}
    However,  once one uses the explicit operators,
 it becomes clear that the use of $W_q$ constraints for $q>3$ is
not practical  since the corresponding constraints are very
complicated. In fact, to the best of our knowledge they were not explicitly
derived for $q>4$.
We now    briefly summarize the $W_q$ constraints approach.\refmark\DVV\
To write down the constraints one
uses a set of $\alpha=1,...,q-1$ scalar fields $\phi_\alpha$
which  are defined by their  mode expansion
$$\pa\phi_\alpha(z) = \sum_{n\in {\cal Z}} a_{n+{\alpha\over q}}
z^{-( n+{\alpha\over q} +1)}.\eqn\mishpa$$
 By identifying   the    modes with the following differential
operators
 $$a_{-n-{\alpha\over q}}={\sqrt{q}\over \lambda}(\nq)t_{n,\alpha}\qquad
a_{n+{\alpha\over q}}={\lambda\over \sqrt{q}}{\pa\over \pa t_{n,\alpha}},
\qquad n\geq 0\eqn \mishan$$
 one derives the following algebra
$$[a_{\nq}, a_{m+{\beta\over q}}]=(\nq)\delta_{n+m+{\alpha+\beta\over
q}}.$$
 One then writes the operators $W^l$ as a polynomials of order $l$ in $\pa
\phi_\alpha$ and imposes the condition that they obey the relevant part of
the $W_q$ algebra. For instance it is straightforward to check that $L_n$
which is defined by
$$ W^2(z)=T(z)=\sum_n L_n z^{-n-2}=
\sum_\alpha \half :\pa\phi_\alpha(z)\pa\phi_{q-\alpha}(z):
+{q^2-1\over 24 q }{1\over z^2}$$  takes the form
$$L_n=\half\sum_{\alpha=1}^{q-1}\sum_{k=-\infty}^\infty a_{k+{\alpha \over q}}
a_{n-1-k+{q-\alpha \over q}}\eqn\mishln$$
 and obey for  $n\geq -1$  the  anomaly free Virasoro algebra.
A similar result for  the $W_n^3$ with $n\geq -2$
  was derived in ref. [\Li].  The constraints on $\tau$, the square root of the
partition function of the perturbed model is then $W^l_n\tau =0$.
Any given correlator can be expressed  as a differential operator acting on
the log of  the partition function as follows
$$\langle P_{k_1,\alpha_1}...P_{k_n,\alpha_n} \rangle =
{\partial \over {\partial t_{k_1,\alpha_1}}}...
{\partial \over {\partial t_{k_n,\alpha_n}}}\log Z=
\sum_g \lambda^{2g-2}\langle P_{k_1,\alpha_1}...P_{k_n,\alpha_n}
\rangle_g$$
 We then express one of the derivatives using
the constraint operators to  derive a recursion relation for the correlator
after setting all the $t_{k,\alpha}$ to their critical values, namely,
$t_{k,\alpha}=0$ apart from $t_{1,1}$.
   For instance
let us calculate the correlation function $\langle P_{0,1}
P_{0,1}P_{0,3}\rangle_0$ in the $(1,4)$ model. For this we use the
constraint
$L_{-1} = {1 \over 2} \sum_{\alpha=1}^3 \sum_{k=-\infty}^\infty a_{k+
{\alpha \over 4}} a_{-2-k+{{4-\alpha} \over 4}} $.
At the critical point  $t_{1,1}=-{4 \over 5}$
 and the only negative $a$'s that survives are $t_{0,1}
\sim a_{-1+{3 \over 4}}$ , $t_{0,3} \sim a_{-1+{1 \over 4}}$ and
$t_{1,1} \sim a_{-2+{3 \over 4}}$. Thus,  we can write
$\{\lambda^{-2}\times {3 \over 4}\times t_{0,1} \times
t_{0,3} + (1+{1 \over
4})\times t_{1,1} \times {\partial \over {\partial t_{0,1}}} + other
\; terms \}\cdot Z=0$
Therefore we can express the derivative of
$Z$ with respect to $t_{0,1}$ as \hfill\break\noindent
$ {\partial \over {\partial t_{0,1}}}Z={3 \over 4}\lambda^{-2}  t_{0,1}
t_{0,3}Z +(other \;terms )Z $
and so
$ \langle P_{0,1} P_{0,1} P_{0,3} \rangle =
{\partial \over {\partial t_{0,3}}}
{\partial \over {\partial t_{0,1}}}
{\partial \over {\partial t_{0,1}}} logZ = {3 \over 4}\lambda^{-2}  $.
Expanding both sides in powers of $\lambda$ we finally get
$ \langle P_{0,1} P_{0,1} P_{0,3} \rangle_0={3 \over 4}$.

 In general,  recursion relations of  products of descendant and primary
operators  for $q=3$  are  composed of  single  and double contact
terms,  single and double surface degeneration terms, degeneration combined
with contact terms and certain Kronecker delta function terms.\refmark\Go

The MR approach\refmark\MR\ is based on inserting a ``complete" set of
states at each degeneration of the surface and summing over all  the possible
degenerations.  Auxiliary  operators  of negative  ghost number were
introduced  in order to define   an adjoint
operator
 on the Hilbert space  of states   $P_i^\dagger =P_{-i}$ where
$i=qk_i+\alpha_i$,
 the metric
$ < P_i P_j>_0=\eta_{ij}=| i|\delta_{(i+j)}$
and the identity
operator  $ I=\sum_{i,j\ne 0(mod q)} |P_i>\eta^{ij}< P_j|$.
It was shown\refmark\MR\  that on the sphere  correlators with
``anti-states" $(i<0)$ could consistently  be set to zero apart from the metric
and the one point function $<P_{-q-1}>_0$.  A recursion relation for any given
correlator on the sphere was then derived using the following procedure.  One
operator is chosen to be the marked operator.  This operator comes in
contact with all possible degenerations of the surface, the number of which is
determined by the primary field  from which the marked operator
descends.  At each degeneration a complete set of states is introduced.   The
recursion relation is then written in terms of a ``degeneration equation"
which states that the sum over all the contacts with degenerations
determined by the marked operator vanishes.  The negative ghost number
operator disappears from the correlators by performing ``effective (multiple)
contacts"  with the marked operator.  The effective contact terms were found
to be
$$\eqalign{\{\{P_{i}\prod_{j=1}^n P_{i_n}\}\}&=
(-1)^{n-1}n!({\alpha \choose n}){\prod_{j=1}^n i_j\over q^n}
 P_{i'} \cr
i'&={i+ \sum_{j=1}^n i_j- n(q+1)} \cr}
\eqn\mishef$$
where $i=kq+\alpha$.
On higher genus Riemann surfaces, one encounter infinities  which follows
from counting ambiguity.  This situation occurs for $q\geq 4$. It
happens on the torus and  even more frequently on surfaces with higher
genus.   Therefore,  an application of the topological procedure for the
general  case at any higher genus Riemann surface  is still missing.

\section{Correlation Functions in the $(1,4)$ Model}
In the previous section correlation functions were shown to follow from the
$W^3$ constraints  as well as from the  MR topological procedure.
 We recall that  these methods are not adequate
already for $q=4$ the former due to regularization problems.
  We now
show that  in fact  using the Viraosro and $W^3$ constraints one can solve
for every correlator in the $(1,4)$ model. We state this result as the
following
theorem.

   {\bf Theorem} -{\it In the $(1,4)$ model every correlation
function at any genus can be computed using the
Virasoro and $W_3$ (\VW) constraints only. }\hfill\break\noindent
{\bf Proof}- By induction on the genus. We first prove that the theorem
holds for $g=0$, then assuming that it holds for certain $g\ge 0$ we show
that it  holds also for $g+1$.
For $g=0$ we  make use of the following lemma: \hfill\break\noindent
{\bf lemma 1}-{\it In the $(1,4)$ model any correlation function at $g=0$
can be computed using the Virasoro and $W_3$
constraints only.}\hfill\break\noindent
{\bf Proof of lemma 1}- By induction on the number $n$ of operators in
the correlation function.  The lemma is proven for $n=3$ and then assuming
it is true for some $n\geq3$ we prove that it holds also for $n+1$.
 For $g=0$, due to ghost number conservation,   any
correlation function includes at least 3 operators.
Correlators of  only $\alpha = 1,2$  primaries   and  of their  descendants
are
obviously determined by the Virasoro and $W^3$ constraint.
 The only non-trivial  three point function  that includes a primary
with $\alpha=3$ or its descendants,  is $\langle P_{0,1} P_{0,1} P_{0,3}
\rangle $.  For this case  the Virasoro constraint is sufficient
( see  the example in
section 2),
 so the claim is proven for
$n=3$.\hfill\break\noindent
The induction hypothesis-{\it  Suppose that the claim in lemma 1 is correct
for correlators that contain $n$ operators, $n \ge 3$.}
 Consider a
correlation function that contains $n+1$ operators. If one of these
operators have $\alpha = 1,2$, the Virasoro or $W^3$ recursion
relation can be used
to calculate the correlator. The result will necessarily contain correlators
with $n$ operators or less. This is  a consequence of a contact that
reduces the number of operators, or  of splitting
the correlator into two or three  correlators  among which the
remaining $n$ operators are distributed. Therefore, all the
 correlation functions that appear after one step of the recursion
fulfill the conditions of the induction assumption. Thus, they
are completely determined by  the Virasoro and $W^3$ recursions.
Consider a correlation function that contains $n+1$ operators, none of
which is with $\alpha = 1,2$. In the $(1,4)$ model the remaining $\alpha$
is 3 and thus the correlation function is of the form $\langle
\prod_{i=1}^{n+1}P_{k_i,3}\rangle_0$.
 Consider the following two correlation functions:
$$\langle P_{0,2} P_{k_1+1,2}\prod_{i=2}^{n+1}P_{k_i,3}
\rangle_0\;\;\;\;\;,
\;\;\;\;\;\langle P_{k_1+1,2} P_{0,2}
\prod_{i=2}^{n+1}P_{k_i,3}\rangle_0$$
Both of them can be computed using the $W^{(3)}$ recursion. For the
first one we get
\harveqn\acorz{ \langle P_{0,2}
P_{k_1+1,2}\prod_{i=2}^{n+1}P_{k_i,3}\rangle_0 =
(2k_1+3)\langle\prod_{i=1}^{n+1}P_{k_i,3}\rangle_0 + A }
where A denotes a set of correlators containing the operator
$P_{k_1+1,2}$
and $n$ other operators (as a consequence of a contact), or $n-1$ other
operators (as a consequence of two contact),
or $n$ operators without $P_{k_1+1,2}$ (again
as a consequence of two contact), or a product of 2 or 3 correlators (as
a consequence of a split) each of them contains $n$ operators
 at the most. All of
these cases can be computed according to the induction
hypothesis.\hfill\break\noindent
For the second correlation function we get
\harveqn\bcorz{ \langle
P_{k_1+1,2}P_{0,2}\prod_{i=1}^{n+1}P_{k_i,3}\rangle_0 =
\langle\prod_{i=1}^{n+1}P_{k_i,3}\rangle_0 + B }
Where the terms in B are calculable in the same way as in A. We can use
now the fact that the correlation
function does not depend on
the order of the operators and to subtract $\acorz$ from $\bcorz$. We
get:
\harveqn\corz{ \langle\prod_{i=1}^{n+1} P_{k_i,3} \rangle_0 =
{1\over{(2k_1+2)}}(B-A)}
Where all the terms on the r.h.s can be computed according to the
induction hypothesis.\hfill\break\noindent
Thus we have proven that a correlation
function containing $n+1$ operators can be computed using the  \VW
recursions only, and this completes the proof of the induction step,
and thus of lemma 1.\hfill\break\noindent

Returning to   the main theorem, we now use again a proof by induction.
We take as our
 induction hypothesis that  the claim of the theorem is correct for genus
$g\geq 0$  and we  would like to show that it holds also for $g+1$.
 To prove the induction step we  use  the following lemma.
\hfill\break\noindent
{\bf Lemma 2}-{\it  If any correlation function up to genus $g$ can be
computed by the Virasoro and $W^{(3)}$ recursion relations only, then any
correlation function at genus $g+1$ can also be computed using these
recursion relations only.}\hfill\break\noindent
{\bf Proof of lemma 2}-  We now apply an  induction on the number $n$ of
operators in the correlation function. For $n=1$, if this operator has
$\alpha=1,2$ then the correlator can be computed using the \VW recursion
relations, having on the r.h.s correlators at genus $g$ or less. If this
operator
has $\alpha=3$, then the correlation function we want to compute is
$\langle P_{k,3}\rangle_{g+1}$, with $k =\half(5g+1)$ due to
ghost-number conservation. This obviously implies that
these  one point functions are non-trivial only on even genus
surfaces.
 Consider the two functions
$$\langle P_{k+1,2} P_{0,2}\rangle_{g+1}\;\;\;\;\;,
\;\;\;\;\;\langle P_{0,2} P_{k+1,2}\rangle_{g+1}$$
As in the case of $g=0$, we can compute them using the $W^{(3)}$
recursion:
\harveqn\acor{\langle P_{k+1,2} P_{0,2} \rangle_{g+1} =  \langle P_{k,3}
\rangle_{g+1} + A }
\harveqn\bcor{ \langle P_{0,2} P_{k+1,2} \rangle_{g+1} = (2k+3) \langle
P_{k,3} \rangle_{g+1} + B }
where   all the terms in A and B are correlation functions at genus
$g$ or less. Subtracting $\acor$ from $\bcor$, we get
\harveqn\cor{ \langle P_{k,3} \rangle_{g+1} = {1 \over {(2k+2)}}(B-A) }
where according to the lemma assumption
all the terms on the r.h.s can be computed using the  \VW
recursion relations.
This completes the
proof for a one point function. On odd genera similar arguments hold for the
 two point function.
We now proceed  by invoking the
 induction hypothesis which states that
{\it any correlation function at genus $g+1$ with
$n$ operators can be computed using the \VW
recursion relations only.}\hfill\break\noindent
Consider a correlation function at genus $g+1$ with $n+1$ operators. If
one of them has $\alpha = 1,2$, then the correlation function can be
computed by the \VW\ recursion relations.
The latter relates the correlator  to a sum of correlators  which  are either
  at genus $g+1$ with $n$ operators or
less,  or $n+1$ operators or less  at genus $g$ or less.
 These correlators are thus completely determined
 according to the induction assumption.
If, on the other hand, all the  operators are of the  $\alpha =3$ type ,namely,
 $\langle
\prod_{i=1}^{n+1} P_{k_i,3} \rangle_{g+1}$,
one repeats the
 treatment  introduced for the  $g=0$ case and derive the same conclusions.
We have thus proven lemma 2 and hence the theorem  that every correlator
of the $(1,4)$ model  can be computed with  the use of the \VW\ constraints
only.
Note that rather than inserting $P_{k_1+1,2}P_{0,2}$ in eqn. \bcorz, one can
use the operators $P_{k_1,2}P_{1,2}$. This  leads to the same conclusions as
above, but cannot be generalized to the higher $(1,q)$ models.
In appendix  A we demonstrate this result by computing several correlation
functions using only the \VW constraints.   The results are in complete
agreement with  the $KdV$ results.\refmark\MRR

\section{Correlation Function For General $(1,q)$ Model}
 We now generalize the result of the previous section to any arbitrary
$(1,q)$ model.  The proof now involves three stages
of induction steps,  one for   the genus,  another one
for  the number of operators and a third one, which was not needed in the
$(1,4)$ case,  for $\alpha= \min_{i=1..N} \{ \alpha_i \}$.

Consider the correlator $\langle \prod_{i=1}^N P_{k_i,\alpha_i}
\rangle_g$  such that
$\alpha \ge 3$. Note that unlike the previous section here we denote the
number of operators in a correlator by $N$ rather than $n$.
 Without a loss of  generality we
 assume $\alpha_1=\alpha$.
To apply the method used for $q=4$ we have to introduce several
additional operators into a correlator that could be related to the
original one. Obviously those operators should be $P_{0,2}$ or
their descendants. In fact, only primaries can do the job since
the contacts which contain only them  vanish and thus their contacts
necessarily involve some of the original operators.
Thus, a situation where we end up with more than $N$ operators is avoided.
 We would like
to get the operator $P_{k_1,\alpha}$ from consecutive contacts of
 the  $P_{0,2}$ primaries  with an operator which is also a descendant of
$\alpha=2$. This requirement determines uniquely the minimal set
of additional operators.  The starting point is, therefore,  the
following two correlators
 $$ \langle  P_{0,2}^{\alpha-2}
P_{k+\alpha-2,2} \prod_{i=2}^N P_{k_i,\alpha_i} \rangle_g \;\;,
\;\; \langle P_{k+\alpha-2,2} P_{0,2}^{\alpha-2} \prod_{i=2}^N
P_{k_i,\alpha_i} \rangle_g \eqn\mishqC$$
As before, we want to subtract the two correlators,
and to extract a recursion relation for the original correlator. After
repeatedly using
the $W^3$ constraint for each $P_{0,2}$ we get $$ C \langle \prod_{i=1}^N
P_{k_i,\alpha_i} \rangle_g =  (C_1-C_2) \langle \prod_{i=1}^N
P_{k_i,\alpha_i} \rangle_g =B-A \eqn\mishC$$
where $C$ is a  numerical coefficient  and $B-A$ is a sum of correlators.
Next we  want to show that
(i) $C\ne 0$ for any allowed values of $N$, $k_i$ and $\alpha_i$ and that
(ii) ,  after exhausting all the
$P_{0,2}$'s, the expression for $B-A$ includes
 terms with $g'<g$ or with $N'<N$ or with $g'=g$ , $N'=N$ but with
$\alpha'<\alpha$ where $\alpha'= min_{i=1..N'} \{ \alpha_i \} $.
 Glancing at the recursion relation derived
from the $W^3$ constraint one can  convince oneself that any  contact
(or any two   contact) reduces the number of operators, while a
split (or  a two split, or contact accompanied by  a split) reduces
either the genus or the number of operators or both. Since each of
the other correlators involves a contact (or  a two contact) between
$P_{0,2}$ and the $\prod_{i=2}^N P_{k_i,\alpha_i}$ or a split where
the $\prod_{i=2}^N P_{k_i,\alpha_i}$ are distributed  among the different
correlators, it is clear that after exhausting all the $P_{0,2}$'s
we must have $g' \le g$ or $ N' \le N$. So let us look at terms
with $N'=N$ and $g'=g$. Clearly at least one of the operators in
the correlator should be different from the original one, otherwise
this term would contribute to C. This  occurs in one
of the successive applications of the $W^3$ constraint due to a
contact or a split, when one of the $P_{k_i,\alpha_i}$ disappear
from the product (possibly with a bunch of $P_{0,2}$) and another
operator appears instead. At least one more new operator must be
formed in order to keep $N'=N$ and it can be formed only from some
of the remaining $P_{0,2}$ and $P_{k+\alpha-j,j}$ where the last
operator is what became of the original $P_{k+\alpha-2,2}$ at this
stage. From ghost number counting it is clear that this new formed
operator has $\alpha' < \alpha$, and that completes the proof of
claim (ii). In order to prove claim (i), we will compute
$C$ explicitly. First we will obtain a general expression
for correlators of the type $\langle P_{0,2}^n P_{k,\alpha}
\rangle_0$ where $n(1-q)+(k-1)q+\alpha-1=-2(q+1)$ and therefore
$k=n-2 \;,\; \alpha=q-n-1 \;,\; 2 \le n <q-1$. The general $W^3$
recursion for correlators of this type is:
$$\eqalign{ \langle P_{0,2}^n
P_{k,\alpha} \rangle_0 &= {1 \over {q^2}} \{
 2 q(kq+\alpha) \langle P_{0,2}^{n-1} P_{k-1,\alpha+1}
\rangle_0
-4 (n-1)(kq+\alpha) \langle P_{0,2}^{n-2}
P_{k-2,\alpha+2} \rangle_0\cr
 &-(kq+\alpha) \sum_{m=2}^{n-3} {{n-1} \choose m} \langle P_{0,2}^m
P_{k_1,\alpha_1} \rangle_0 \langle P_{0,2}^{n-m-1} P_{k_2,\alpha_2}
\rangle_0\cr}\eqn\mishrec $$
Let us now prove by induction that the outcome of this recursion is given by
$$\langle P_{0,2}^n P_{k,\alpha} \rangle_0 = {{2^n \prod_{i=1}^{n-1}
(iq-n-1)} \over q^{n-1}} \delta_{k,n-2} \delta_{\alpha,q-n-1}=
{2^n \Gamma(n
-{n+1\over q})\over \Gamma(1-{n+1\over q})}\delta_{k,n-2}
\delta_{\alpha,q-n-1}\eqn\mishre$$
 It is straightforward to check that  the values of the
correlators for the  first few cases agree with that expression.
 The results are the following
 $$\langle P_{0,2}^2 P_{0,q-3} \rangle_0 = {{4(q-3)} \over q} \;\;\;\;
\;\;\;\;n=2$$
$$\langle P_{0,2}^3 P_{1,q-4} \rangle_0 = {{8(2q-4)(q-4)} \over q^2}
\;\;\;\;\;\;\;\; n=3$$
$$ \langle P_{0,2}^4 P_{2,q-5} \rangle_0 = {{16(3q-5)(2q-5)(q-5)} \over
q^3} \;\;\;\;\;\;\;\; n=4$$
Assuming now that \mishre\ holds for
 $n-1$  and  then using the recursion relation \mishrec\ one finds that
\mishre\
is correct also for $n$, provided that  the following is an
identity
$$  \eqalign{\prod_{i=1}^{n-2} (iq-n-1)=&\prod_{i=1}^{n-2} (iq-n) -{1\over 2}
\sum_{m=1}^{n-2} {{n-1} \choose m} \prod_{i=1}^{m-1} (iq-m-1)
\prod_{j=1}^{n-m-2}
(jq-n+m) \cr
=&-{1\over 2}\sum_{m=0}^{n-1} {{n-1} \choose m} \prod_{i=1}^{m-1} (iq-m-1)
\prod_{j=1}^{n-m-2}
(jq-n+m)\cr}
  \eqn\mishId$$
where in the second line we use the extrapolation of
$\prod_{i=n_i}^{n_f}={\prod_{i=n_j}^{n_f}\over \prod_{i=n_j}^{n_i-1}}$
 to the case that $n_i>n_f$.
It is interesting to note that for the unphysical case of $q=0$ this relation
reduces
to the   identity of Abel\refmark\ABEL
  $$2(n+1)^{(n-2)} = \sum_{m=0}^{n-1}
{n-1\choose m}(m+1)^{(m-1)}(n-m)^{(n-m-2)}\eqn\mishIqz$$
which  counts the number of trees formed from $n+1$ points where two
given points are connected by a link.
 The  identity \mishId\  will  be shown to be a special case of  a more
general identity which is proven in
 appendix B.

Equipped with the expression for the $\langle P_{0,2}^n
P_{k,\alpha} \rangle_0 $ we now proceed to
 write down a  recursion  relation for  a correlator of the type $\langle
P_{0,2}^n P_{k+n,\alpha-n} \prod_{i=2}^N P_{k_i,\alpha_i} \rangle_g $
and consider only terms that contribute to $C_1$
$$ \eqalign{&\langle P_{0,2}^n P_{k+n,\alpha-n} \prod_{i=2}^N
P_{k_i,\alpha_i} \rangle_g =
 {1 \over q^2} \{  2q[(k+n)q+\alpha-n] \langle P_{0,2}^{n-1}
P_{(k,\alpha)(n-1)} \prod_{i=2}^N P_{k_i,\alpha_i} \rangle_g \cr
&-4(n-1)[(k+n)q+\alpha-n] \langle P_{0,2}^{n-2}
P_{(k,\alpha)(n-1)} \prod_{i=2}^N P_{k_i,\alpha_i} \rangle_g
\cr &- 2[(k+n)q+\alpha-n] \sum_{m=2}^{n-1} {{n-1} \choose m}
\langle P_{0,2}^m P_{\tilde k, \tilde \alpha} \rangle_0 \langle
P_{0,2}^{n-1-m} P_{(k,\alpha)(n-m-1)} \prod_{i=2}^N
P_{k_i,\alpha_i} \rangle_g +\Delta_1 \cr}\eqn\mishpoly$$
where  $P_{(k,\alpha)(l)}$ means $P_{k+l,\alpha-l}$ and
$\Delta_1$ stands for  terms that do not contribute to $C_1$.

Once again we make use of an induction procedure and a
polynomial identity to prove a
result. This time it is
  the general expression for $C_1$, the contribution of this correlator to
$C$, which takes the form
$$ C_1=({2 \over q})^n \prod_{i=1}^n [(k+i)q+\alpha-n] $$
The cases of $n=1$ and $n=2$ can be easily shown to obey this rule.
When one  assumes it holds for  $n-1$ and uses the
recursion \mishpoly\ one proves the result for $n$ if the following expression
is an
identity.
$$ \eqalign{&\prod_{i=1}^{n-1} [(k+i)q+\alpha-n+1] - (n-1) \prod_{i=1}^{n-2}
[(k+i)q+\alpha-n+2] \cr
&-\sum_{m=2}^{n-1} {{n-1} \choose m} \prod_{i=1}^{m-1} (iq-m-1)
\prod_{i=1}^{n-1-m} [(k+i)q+\alpha-n+1-m] \cr
&=-\sum_{m=0}^{n-1} {{n-1} \choose m} \prod_{i=1}^{m-1} (iq-m-1)
\prod_{i=1}^{n-1-m} [(k+i)q+\alpha-n+1-m] \cr
&= \prod_{i=1}^{n-1} [(k+i)q+\alpha-n] \cr}\eqn\mishIdd$$
The proof of this identity is given
in appendix B. Note that for the special
value $\alpha +2 +kq=0$ this identity turns into the previous one given in
\mishId.

The next step is obviously the computation of  the contribution to $C$ from
 $\langle P_{k+n,\alpha-n} P_{0,2}^n \prod_{i=2}^N P_{k_i,\alpha_i}
\rangle_g $ .
The recursion relation in this case reads

$$ \eqalign{&\langle P_{k+n,\alpha-n} P_{0,2}^n \prod_{i=2}^N P_{k_i,\alpha_i}
\rangle_g =
 {1 \over q^2} \{4qn \langle P_{0,2}^{n-1} P_{(k,\alpha )(n-1)}
\prod_{i=2}^N P_{k_i,\alpha_i} \rangle_g \cr
 &+2q \sum_{m=2}^n {n \choose m} \langle P_{0,2}^m P_{\tilde k, \tilde
\alpha} \rangle_0 \langle P_{0,2}^{n-m} P_{(k,\alpha )(n-m)}
\prod_{i=2}^N P_{k_i,\alpha_i} \rangle_g
 -8 {n \choose 2} \langle P_{0,2}^{n-2} P_{(k,\alpha )(n-2)}
\prod_{i=2}^N P_{k_i,\alpha_i} \rangle_g \cr
&-4n \sum_{m=2}^{n-1} {{n-1} \choose m} \langle P_{0,2}^m P_{\tilde k,
\tilde \alpha} \rangle_0 \langle P_{0,2}^{n-m-1}
P_{(k,\alpha )(n-m-1)} \prod_{i=2}^N P_{k_i,\alpha_i} \rangle_g \cr
 &-3 \sum_{a=2}^{n-2} \sum_{b=2}^{n-a} {n \choose a} {{n-a} \choose b}
\langle P_{0,2}^a P_{\tilde k_1, \tilde \alpha_1} \rangle_0 \langle
P_{0,2}^b P_{\tilde k_2, \tilde \alpha_2} \rangle_0
 \langle P_{0,2}^{n-a-b} P_{(k,\alpha )(n-a-b)} \prod_{i=2}^N
P_{k_i,\alpha_i} \rangle_g + \Delta_2\}  \cr}\eqn\mishidd$$
with $\Delta_2$ denotes terms that do not contribute to $C_2$.
And so the contribution to $C_2$ is given by
$$ ( {2 \over q} )^n \bigl \{ 2n \prod_{i=1}^{n-1} [(k+i)q+\alpha-n+1]
+2 \sum_{m=2}^n {n \choose m} \prod_{i=1}^{m-1} (iq-m-1)
\prod_{i=1}^{n-m} [(k+i)q+\alpha-n+m] $$
$$ -2 {n \choose 2} \prod_{i=1}^{n-2} [(k+i)q+\alpha-n+2] -2n
\sum_{m=2}^{n-1} {{n-1} \choose m} \prod_{i=1}^{m-1} (iq-m-1)
\prod_{i=1}^{n-m-1} [(k+i)q+\alpha-n+m+1] $$
$$ - \sum_{a=2}^{n-2} \sum_{b=2}^{n-a} {n \choose a} {{n-a} \choose b}
\prod_{i=1}^{a-1} (iq-a-1) \prod_{i=1}^{b-1} (iq-b-1)
\prod_{i=1}^{n-a-b} [(k+i)q+\alpha-n+a+b] \bigr \} $$
The sums of  products and  triple products in this expression
can be simplified using
eqn.\mishIdd.
  Substituting
    $n=\alpha-2$ into $C_1-C_2$
 we finally   get
$$ C = C_1-C_2 = 2^{\alpha-2} \prod_{i=1}^{\alpha-2} (k+i). \eqn\mishC $$
This completes the proof of claim (i) since obviously $C$ does not vanish.
Having the explicit expression for $C$ is obviously also useful for the full
determination of the correlators.  Note that  $C$ is independent of $q$.
In particular eqn. \corz\  is a special case of \mishC\ for $\alpha=2$.
 The procedure just described to write down a recursion
relation for $\langle \prod_{i=1}^N P_{k_i,\alpha_i} \rangle_g$  constitutes a
proof to the following lemma.

 {\bf Lemma 3:} {\it  In any $(1,q)$ model a recursion
relation can be written for
$\langle \prod_{i=1}^n P_{k_i,\alpha_i} \rangle_g$ with $\alpha \geq 3$
, $\alpha=\min\{\alpha_i\}$ using only the $W^{(3)}$ recursion relations
so that all the terms on the right hand side will be of the form
$\langle \prod_{j=1}^{n'} P_{k'_j,\alpha'_j} \rangle_{g'}$ with $g'<g$
or $n'<n$ or $g'=g$ , $n'=n$ and $\alpha'=\alpha-1$ where
$\alpha'=\min\{\alpha'_j\}$. } \hfill\break\noindent
Using Lemma 3  and the analogs of lemmas 1, 2, proven in the previous
section for the $(1,4)$ model,
 we can now generalize the theorem that any correlator can be
computed using \VW constraints proven  above
 for the $(1,4)$ model to any $q$.
As for Lemma 1, the $g=0$ case,   the smallest number of  operators in any
non-trivial correlator  is still three.   But  it is not necessary
that  all $\alpha=1,2$  if all the
operators are primaries and  $q\geq 8$.
In the latter case we have to use the commutator construction given above
in lemma 3.
 Once the three point function is computed using the
\VW constraints, lemma 1 follows just as for the $q=4$ models.
 \section{ Summary and  Discussion}
Topological properties of  certain moduli spaces can be easily
 written down in terms
of correlators of  topological quantum field theories.  However,
 the field theory realization does not  shed much light on the explicit
evaluation  of  these  topological
characteristics  since  only in a limited number of cases the relevant
correlators could be computed.
 One such example is the moduli space of punctured Riemann surfaces which
corresponds to the theory of pure topological gravity, the topological
$(1,2)$ minimal model.  Expectation values of ``physical operators"  in this
model were determined via
  recursion relations which originally\refmark\wtrr\
  led to a solution on the sphere and later turned  into a  full solution
with the introduction of the contact algebra concept.\refmark\VV\  A
different, even though not  unrelated approach, to this model was
introduced via the Virasoro constraints. The equivalence of the two
approaches  was proven  in ref. [\MR]
 together with a generalization to the $(1,q)$ models.   The  main advantage
of the latter  approach  was the introduction  of a prescription for
 recursion relations  which was much simpler  to use than
 the method of  $W_q$ constraints . In fact
  these constraints were never written down for
$q>4$ and already for $q\leq 4$ the resulting expressions are  very
cumbersome.    The MR  approach failed short in computing correlators for
$q>3$  on Riemann  surfaces at genus greater than zero. The  reason for that
was the lack of a regularization scheme to handle
  divergences that appeard  at $q>3$.

The question that was investigated
in the present  work was whether one could   solve  for the correlators of
higher $q$ models at any genus using only
 the  part of  the MR technique which is otherwise
needed for $q=2,3$, or equivalently using only the \VW\ constraints.
Recall that
normal ordering in the Virasoro case ($q=2$)  is well known and the $W^3$
case is free from infinities. Indeed,
we proved by induction that  the  \VW are sufficient for a complete solution
of the models. The fact that it is unnecessary  to use higher $W_q$
information was proven  here only within the framework of the
observables of the topological $(1,q)$ models. The implications to other
domains where $W_q$ algebras are involved,  has to be further
investigated.  One feature which played an important role in the  present
derivation is obviously the interchange of  the order of operators   inside
correlation functions.

There are certain open questions and related topics that are still awaiting
further investigation. Among them one can find: (i)  The search for
a general structure  for classes of correlators   like for
instance eqn. \mishre\ and their interpretation as intersection numbers or
similar topological properties of the corresponding moduli spaces.
(ii) A related question is  whether  there are additional independent
polynomial identities.  Certain generalization of
those presented in this work  have been  already
discovered and they will be presented in a future publication.
(iii)  Another challenge is  to transcribe
  our results   directly in terms of  the
corresponding KdV flows.

 \ack{ We would like to thank  D. Montano and G. Rivlis for  useful
discussions and for letting us use their computer programs.
We thank S. Yankielowicz for many valuable discussions and for reading the
manuscript.  We thank O.Kenet for the proof of eqn.\mishIdd.
 We would like also to thank P. Di Francesco and A. Morozov   for
useful conversations. }

     \Appendix{ A}
\centerline{\bf  Computation of Correlators  of the $(1,4)$ model
using the \VW recursion relations}
The algorithm of computing   correlators,  $\langle
\prod_i^N P_{k_i,3}  \rangle_g$
   that  otherwise would
require  the use of $W_4$ constraint,   using only the \VW ones is
demonstrated here in three examples at genus $g=0,1,2$

$(i)$. As the first example we compute is $\langle P_{0,3}^5
\rangle_0$ . So we look at $\langle P_{0,2} P_{1,2} P_{0,3}^4 \rangle_0$
and $\langle P_{1,2} P_{0,2} P_{0,3}^4 \rangle_0$ . For the first one we
use the $W_{-2}^{(3)}$ constraint:
$$ \sum_k \sum_l a_{k+{1 \over 4}} a_{l+{1 \over 4}} a_{-3-k-l+{1 \over
2}} + \sum_k \sum_l a_{k+{3 \over 4}} a_{l+{3 \over 4}}
a_{-4-k-l+{1 \over  2}} $$
The negative $a$'s that survives at the critical point are
$a_{-2+{1 \over 2}} \sim t_{1,2}$ ,
$a_{-1+{1 \over 4}} \sim t_{0,3}$ ,
$a_{-1+{1 \over 2}} \sim t_{0,2}$ , and
$a_{-2+{3 \over 4}} \sim t_{1,1}$ . So the relevant terms in the
constraint are
$$ 2 \times {2 \over \lambda} \times {3 \over 2} \times t_{1,2} \times
{3 \over 4} \times t_{0,3} \times {\partial \over {\partial t_{0,1}}} +
{2 \over \lambda} \times [(1+{1 \over 4}) \times t_{1,1}]^2 \times
{\partial \over {\partial t_{0,2}}} + 2 \times {2 \over \lambda} \times
{3 \over 2} \times t_{1,2} \times (1+ {1 \over 4}) \times t_{1,1} \times
{\partial \over {\partial t_{0,3}}}$$
Translating this to correlators we get
$$\langle P_{0,2} P_{1,2} P_{0,3}^4 \rangle_0 = 3 \langle P_{0,3}^5
\rangle_0 - 9 \langle P_{0,1} P_{0,3}^3 \rangle_0$$
Similarly we get for the second correlator
$$\langle P_{1,2} P_{0,2} P_{0,3}^4 \rangle_0  = - {3^3 \over 4} \langle
P_{0,2}^2 P_{0,3}^2 \rangle_0 -3 \langle P_{0,1} P_{0,3}^3 \rangle_0 +
\langle P_{0,3}^5 \rangle_0$$
combining these results we get
$$ \langle P_{0,3}^5 \rangle_0 = {1 \over 2} ( 6 \langle P_{0,1}
P_{0,3}^3 \rangle_0 - {3^3 \over 4} \langle P_{0,2}^2 P_{0,3}^2
\rangle_0 )$$
Inserting      the following values, which
can be computed directly from the constraints
 $$ \langle P_{0,1} P_{0,3}^3 \rangle_0 = 0 \;\;\;\;,\;\;\;\; \langle
P_{0,2}^2 P_{0,3}^2 \rangle_0 = -{9 \over 4}$$
We finally get:
$$\langle P_{0,3}^5 \rangle_0 = {243 \over 32} = \bigl ( {3 \over 2}
\bigr )^5$$
 $(ii)$ $\langle
P_{1,3} P_{0,3} \rangle_1$.

We have
$$\langle P_{0,2} P_{2,2} P_{0,3} \rangle_1 = 2 \times{5 \over 2} \times
\langle P_{1,3} P_{0,3} \rangle_1 - 2 \times {5 \over 2} \times {3 \over
4} \times \langle P_{1,1} \rangle_1 -{5 \over 2} \times {1 \over 4}
\times \langle P_{0,1} P_{0,1} P_{0,3}\rangle_0 $$
And also
$$\langle P_{2,2} P_{0,2} P_{0,3} \rangle_1 = 2 \times {1 \over 2}
\times \langle P_{1,3} P_{0,3} \rangle_1 - 2 \times {1 \over 2} \times
{3 \over 4} \times \langle P_{1,1} \rangle_1 - {1 \over 2} \times {1
\over 4} \times \langle P_{0,1} P_{0,1} P_{0,3} \rangle_0 $$
$$ + 2 \times {1 \over 4} \times \langle P_{0,2} P_{0,2} P_{0,3} P_{0,3}
\rangle_0 -2 \times {1 \over 4} \times {3 \over 4} \times \langle
P_{0,2} P_{0,2} P_{0,1} \rangle_0 $$
And thus
$$ \langle P_{1,3} P_{0,3} \rangle_1 = {1 \over 4}
\Bigl(3 \langle P_{1,1}
\rangle_1 + {1 \over 2} \langle P_{0,1} P_{0,1} P_{0,3} \rangle_0$$
$$ + {1
\over 2} \langle P_{0,2} P_{0,2} P_{0,3} P_{0,3} \rangle_0 - {3 \over 8}
\langle P_{0,2} P_{0,2} P_{0,1} \rangle_0 \Bigr) $$
And if we substitute the values(computed directly from the $W_3$
constraints):
$$\langle P_{1,1} \rangle_1 = {5 \over 32} \;\;\;\;\;,\;\;\;\;\; \langle
P_{0,1} P_{0,1} P_{0,3} \rangle_0 = {3 \over 4}$$
$$\langle P_{0,2} P_{0,2} P_{0,3} P_{0,3} \rangle_0 = -{9 \over
4}\;\;\;\;\;,\;\;\;\;\; \langle P_{0,2} P_{0,2} P_{0,1} \rangle_0 = 1$$
We get:
$$\langle P_{1,3} P_{0,3} \rangle_1 = - {21 \over 128} $$
$(iii)$ Our third example is $\langle P_{3,3} \rangle_2$. From the
constraints we get:
$$\langle P_{1,2} P_{3,2} \rangle_2 =-{175 \over 1024} - {7 \over 4}
\langle P_{0,1} P_{2,1} \rangle_1 - {7 \over 4} \langle P_{0,3} P_{1,3}
\rangle_1 + 7 \langle P_{3,3} \rangle_2$$
$$ \langle P_{3,2} P_{1,2} \rangle_2 = - {1 \over 16} \langle P_{0,1}^2
P_{0,2} P_{1,2} \rangle_0 -{75 \over 1024} - {3 \over 4} \langle P_{0,1}
P_{2,1} \rangle_1$$
$$+ 3 \langle P_{3,3} \rangle_2 + {1 \over 2} \langle P_{0,3} P_{1,2}^2
\rangle_1 + {1 \over 2} \langle P_{1,2} P_{0,2} P_{1,3} \rangle_1 -{3
\over 4} \langle P_{0,3} P_{1,3} \rangle_1$$
And so we get:
$$\langle P_{3,3} \rangle_2 = {1 \over 4} ( {25 \over 256} + \langle
P_{0,1} P_{2,1} \rangle_1 + \langle P_{0,3} P_{1,3} \rangle_1 - {1 \over
16} \langle P_{0,1}^2 P_{0,2} P_{1,2} \rangle_0$$
$$ + {1 \over 2} \langle P_{0,3} P_{1,2}^2 \rangle_1 + {1 \over 2}
\langle P_{1,3} P_{0,2} P_{1,2} \rangle_1 ) = - {263 \over 1024}$$
\Appendix{ B}
\centerline{ \bf  Polynomial identities}

The steps toward the proof of Lemma 3  included  the use of the identities
$$ -{1\over 2} \sum_{m=0}^{n-1} {{n-1}
\choose m} \prod_{i=1}^{m-1} (iq-m-1) \prod_{j=1}^{n-m-2}
(jq-n+m)
 = \prod_{i=1}^{n-2} (iq-n-1) \eqn\mishId$$
and
$$ \eqalign{
&-\sum_{m=0}^{n-1} {{n-1} \choose m} \prod_{i=1}^{m-1} (iq-m-1)
\prod_{i=1}^{n-1-m} [(k+i)q+\alpha-n+1+m] \cr
&= \prod_{i=1}^{n-1} [(k+i)q+\alpha-n]. \cr}\eqn\mishIdd$$
As stated in section 4. identity \mishId\
  is a special case   of \mishIdd\ at $\alpha+2 + kq=0$,
  and Abel's identity  (18)
  is the $q=0$ case of \mishId. We thus present here the proof of
eqn.\mishIdd.

{\bf Proof of identity \mishIdd\ }\footnote{*}{ The identity was proven by
O.Kenet.}

A product of the form $\prod_{i=1}^{n} [ iq-\gamma]$ can be represented
in the following form
$$\prod_{i=1}^{n} [iq+\gamma] x^{{\gamma\over q}-(n+1)}=
(-q)^n({d\over dx})^n [x^{{\gamma\over q}-1}]\equiv f(x,n,q,\gamma)
\eqn\mishder$$
In terms of this representation \mishIdd\ takes the form
$$({d\over dx})^{n-1} [x^{{n-\beta\over
q}-1}]=-\sum_{m=0}^{n-1} {{n-1} \choose m} ({d\over dx})^{m-1}
[x^{{m+1\over q}-1}]({d\over dx})^{n-m-1} [x^{{n-\beta-m-1\over
q}-1}]\eqn\mishderi$$
where $\beta=\alpha+kq$

It is easy to check that the identity holds for $n=1,2$. We now assume that
it holds for $n$ and we examine now the case of $n+1$.
In the RHS that implies an additional differentiation with respect to $x$ and
thus we now take the derivative of the RHS
$$\eqalign{RHS=&-\sum_{m=0}^{n-1} {{n-1} \choose m} ({d\over dx})^{m}
[x^{{m+1\over q}-1}]({d\over dx})^{n-m-1} [x^{{n-\beta-m-1\over
q}-1}]\cr
&-\sum_{m=0}^{n-1} {{n-1} \choose m-1} ({d\over dx})^{m-1}
[x^{{m+1\over q}-1}]({d\over dx})^{n-m} [x^{{n-\beta-m-1\over
q}-1}]\cr}\eqn\mishpr$$
We now replace the summation index  $m$ in the first term  as follows
$m\rightarrow n-m $ and we choose a particular value for $\beta$,
$\beta=-2$. The two terms are now identical apart from the combinatorial
factor and the summation range. The latter is easily fixed  and  using
${{n-1} \choose m-1}+{{n-1} \choose m} ={{n} \choose m}$ one gets
$$RHS=-\sum_{m=0}^{n} {{n} \choose m} ({d\over dx})^{m}
[x^{{m+1\over q}-1}]({d\over dx})^{n-m} [x^{{n-m+1\over
q}-1}]\eqn\mishpb$$
which is exactly the RHS of the identity \mishIdd\ for $n+1$ at the particular
value of $\beta=\beta_0=-1$. Differentiating the LHS at $\beta=-2$ also takes
the form of the LHS  at $n+1$ and $\beta=\beta_0=-1$.  The identity was
thus proven for the particular value $\beta_0$.

To prove the  identity for any value of $\beta$ we now show that if it holds
for some particular value $\beta_0$ it holds also for $\beta_0-q$.  By
repeating this process  the identity which is a polynomial in beta is shown to
hold in infinitely many points and thus it should be an exact identity.
It is easy to realize that
$$f(x,n,q,\gamma-q)= xf(x,n,q,\gamma) +nf(x,n-1,q,\gamma)$$
 and thus inserting it into  eqn.\mishIdd\ with the use of the assumption that
holds for $n-1$ for any $\gamma$ we get that the identity holds for the
shifted $\beta$.
This completes the proof of the identity.
\endpage
\refout
 \end
\bye